# Bohmian mechanics versus Madelung quantum hydrodynamics

Roumen Tsekov
Department of Physical Chemistry, University of Sofia, 1164 Sofia, Bulgaria

It is shown that the Bohmian mechanics and the Madelung quantum hydrodynamics are different theories and the latter is a better ontological interpretation of quantum mechanics. A new stochastic interpretation of quantum mechanics is proposed, which is the background of the Madelung quantum hydrodynamics. Its relation to the complex mechanics is also explored. A new complex hydrodynamics is proposed, which eliminates completely the Bohm quantum potential. It describes the quantum evolution of the probability density by a convective diffusion with imaginary transport coefficients.

The Copenhagen interpretation of quantum mechanics is guilty for the quantum mystery and many strange phenomena such as the Schrödinger cat, parallel quantum and classical worlds, wave-particle duality, decoherence, etc. Many scientists have tried, however, to put the quantum mechanics back on ontological foundations. For instance, Bohm [1] proposed an alternative interpretation of quantum mechanics, which is able to overcome some puzzles of the Copenhagen interpretation. He developed further the de Broglie pilot-wave theory and, for this reason, the Bohmian mechanics is also known as the de Broglie-Bohm theory. At the time of inception of quantum mechanics Madelung [2] has demonstrated that the Schrödinger equation can be transformed in hydrodynamic form. This so-called Madelung quantum hydrodynamics is a less elaborated theory and usually considered as a precursor of the Bohmian mechanics. The scope of the present paper is to show that these two theories are different and the Madelung hydrodynamics is a better interpretation of quantum mechanics than the Bohmian mechanics. A stochastic interpretation is also developed, which is the background of the Madelung quantum hydrodynamics [3].

The evolution of the wave function $\psi$ of a quantum mechanical system consisting of $N$ particles is described by the Schrödinger equation

$$i\hbar \partial_t \psi = (-\hbar^2 \nabla^2 / 2m + U)\psi \qquad (1)$$

where $\nabla$ is a $3N$-dimensional nabla operator and $U$ is a potential. The complex wave function can be presented generally in the polar form

$$\psi = \sqrt{\rho}\exp(iS/\hbar) \qquad (2)$$

where $\rho = |\psi|^2$ is the $N$-particles distribution density and $S/\hbar$ is the wave function phase. Introducing Eq. (2) in the Schrödinger equation (1) results in two equations

$$\partial_t \rho = -\nabla \cdot (\rho \nabla S / m) \tag{3}$$

$$\partial_t S + (\nabla S)^2 / 2m + U + Q = 0 \tag{4}$$

where $Q \equiv -\hbar^2 \nabla^2 \sqrt{\rho} / 2m\sqrt{\rho}$ is the so-called quantum potential. Bohm [1] has noticed that in the classical limit $Q$ vanishes and Eq. (4) reduces to the Hamilton-Jacobi equation. For this reason, he suggested that $S$ is the mechanical action, which is related to the real velocities of the particles via the relation [1]

$$\dot{R} = \nabla S / m \tag{5}$$

Here $R$ is the $3N$-dimensional vector of the particles coordinates. Using expression (5) and Eq. (4) one can easily derive a quantum Newtonian equation [1]

$$m\ddot{R} = -\nabla(U + Q) \tag{6}$$

This equation hints already inconsistency of the Bohmian mechanics since the particles trajectories depend via $Q$ on the probability density to find the particles at their places, while the logic of a statistical mechanics is just the opposite. Some authors try to resolve this philosophical discrepancy by reinterpretation of the quantum probability [4]. We consider the de Broglie-Bohm theory as something like a mean-field approximation of the real quantum dynamics, which is stochastically driven by the vacuum fluctuations.

Bohm paid much attention to Eq. (4) and less concern about Eq. (3). It is easy to check that the solution of the system of Eqs. (3) and (5) is the probability density $\rho = \delta(r - R)$, which is typical for a deterministic motion described particularly in the Bohmian mechanics via Eq. (6). This Bohmian distribution cannot describe, however, the probability density from the quantum mechanics since its dispersion is always zero. Hence, the Bohmian mechanics contradicts to the quantum mechanics. In general any distribution density can be presented in the form $\rho = <\delta(r - R)>$, where the brackets indicate statistical average over the realizations of the particles trajectories. The stochasticity of $R$ could originate either from unknown initial conditions, how Bohm proposed [1], or from some inherent fluctuations [3]. Taking a time derivative of this expression leads straightforward to the continuity equation

$$\partial_t \rho = -\nabla \cdot (\rho V) \tag{7}$$

where the $3N$-dimensional velocity-vector is given by $V \equiv <\dot{R}\delta(r-R)>/\rho$. This hydrodynamic-like velocity obviously is not simply the particles velocity but an averaged product representing the flow in the probability space. As seen, Eq. (7) is general and not specifically related to the quantum mechanics. Comparing now Eq. (3) and Eq. (7) one concludes that $S$ is the hydrodynamic-like velocity potential, not the mechanical action as suggested by Bohm. Hence, the correct alternative of Eq. (5) reads

$$V = \nabla S / m \tag{8}$$

Let us check now if Eq. (4) can be also explained in this scheme. Using Eq. (8) it can be transformed easily to a macroscopic force balance for the hydrodynamic-like velocity

$$m\partial_t V + mV \cdot \nabla V = -\nabla(U + Q) \tag{9}$$

Therefore, Eq. (4) is not a quantum Hamilton-Jacobi equation as Bohm suggested. The system of Eqs. (7) and (9) was proposed for a single particle by Madelung [2] first and is known in the Science as the Madelung quantum hydrodynamics. Now the probability density is not driving the individual particles via $Q$ but their hydrodynamic-like velocity, which is similar to the thermal diffusion. In the latter case the driving force is the gradient of the local Boltzmann entropy, while the quantum potential is proportional to the local Fisher entropy [5]. The latter accounts for the quality of measurements. Thus, according to the Cramér–Rao bound, the Heisenberg Uncertainty principle is equivalent to a standard inequality for the efficiency of measurements.

In contrast to the Bohmian mechanics, the Madelung quantum hydrodynamics describes only the averaged statistical characteristics $\rho$ and $V$ but not the particles trajectory $R$. Since the latter is stochastic and the vacuum is a non-dissipative environment one can propose the following stochastic quantum Newtonian equation

$$m\ddot{R} = -\nabla U + f_Q \tag{10}$$

where $f_Q$ is a random force originating from some vacuum fluctuations [3]. Its average value is zero to satisfy the Ehrenfest theorem. According to the modern physics all the interaction in the world are mediated by virtual particles and Eq. (10) states that the latter behave randomly as well. The quantum potential is the macroscopic image of the microscopic force $f_Q$. Hence, in a mean-field approach one can replace $f_Q$ by $-\nabla Q$ to get Eq. (6). The phase-space probability density can be generally presented via $W \equiv <\delta(p-m\dot{R})\delta(r-R)>$. Differentiating $W$ in time and expressing the particles acceleration from Eq. (10) yields

$$\partial_t W + p \cdot \nabla W / m - \nabla U \cdot \partial_p W + \partial_p \cdot < f_Q \delta(p-m\dot{R})\delta(r-R) > = 0 \qquad (11)$$

In the classical limit the last quantum term vanishes and Eq. (11) reduces to the Liouville equation. If one assumes now that the unspecified quantum force term is given by

$$< f_Q \delta(p-m\dot{R})\delta(r-R) > = -\sum_{k=1}^{\infty} \frac{(i\hbar/2)^{2k}}{(2k+1)!} \nabla^{2k+1} U \cdot \partial_p^{2k} W \qquad (12)$$

Eq. (11) becomes the Wigner-Liouville equation [6], from which the Madelung quantum hydrodynamics can be deduced straightforward [7]. According to Eq. (12) the quantum stochastic force is not correlated to the particle position since $< f_Q \delta(r-R) > = 0$. One of the advantages of the Bohmian mechanics is a demonstration of quantum non-locality, which is due to the fact that $Q$ is a function of the positions of all the particles in the system. At a first look Eq. (10) seems local and one could pretend that it violates the Bell theorem. However, the stochastic forces acting on different particles are obviously correlated since the same quantum potential appears in the non-local Madelung hydrodynamics as well. Hence, the present stochastic interpretation does not only reproduce the quantum non-locality but shows the physical reason for the entanglement: the spatial correlations of the vacuum fluctuations.

An alternative way to describe the present stochastic quantum dynamics is the complex mechanics [8, 9]. According to this theory quantum particles obey also the Newtonian equation $m\ddot{Z} = -\nabla U(Z)$ but their coordinates $Z(t)$ are complex functions of time. The real part $R = Z_{\text{Re}}$ represents the observable physical trajectories. Since the initial value of the metaphysical imaginary part $Z_{\text{Im}}$ is unknown, the complex mechanics description possesses a stochastic character. For instance, the effect of vacuum fluctuations can be attributed to $Z_{\text{Im}}$. It is easy to show that the phase-space probability density $W$ obeys in complex mechanics the Liouville equation

$$\partial_t W + p \cdot \nabla W / m - \partial_p \cdot < \text{Re}[\nabla U(r+iZ_{\text{Im}})]\delta(p-m\dot{R})\delta(r-R) > = 0 \qquad (13)$$

Expanding now the potential energy $U$ in a power series of the imaginary part $Z_{\text{Im}}$ one can rewrite Eq. (13) in the more decisive form

$$\partial_t W + p \cdot \nabla W / m - \partial_p \cdot \sum_{k=0}^{\infty} \frac{1}{2k!} \nabla^{2k+1} U \cdot < (iZ_{\text{Im}})^{2k} \delta(p-m\dot{R})\delta(r-R) > = 0 \qquad (14)$$

Thus, an alternative expression for the fluctuation force term from Eq. (11) reads

$$<f_Q \delta(p-m\dot{R})\delta(r-R)> = -\sum_{k=1}^{\infty} \frac{1}{2k!} \nabla^{2k+1} U \cdot <(iZ_{Im})^{2k} \delta(p-m\dot{R})\delta(r-R)> \quad (15)$$

The structure of this equation is similar to Eq. (12). Hence, by proper modeling of the statistical properties of $Z_{Im}$ one could derive the Wigner-Liouville equation, i.e. the quantum mechanics.

An interesting alternative of the Madelung quantum hydrodynamics is the complex hydrodynamics with an irrotational complex hydrodynamic velocity

$$\omega \equiv i\hbar \nabla \ln \bar{\psi} / m = V + i\hbar \nabla \ln \rho / 2m \quad (16)$$

introduced via the quantum momentum operator [8, 9]. Substituting this expression, the continuity Eq. (7) changes to a complex convective diffusion equation

$$\partial_t \rho + \nabla \cdot (\rho \omega) = D \nabla^2 \rho \quad (17)$$

with imaginary diffusion coefficient $D \equiv i\hbar / 2m$. The latter is well-known in the physical literature [10, 11], since the Schrödinger equation for a free particle $\partial_t \psi = D \nabla^2 \psi$ formally coincides with a diffusion equation. Using further Eq. (16), the Madelung hydrodynamic force balance (9) acquires the form of a complex Navier-Stokes equation

$$\partial_t \omega + \omega \cdot \nabla \omega = -\nabla U / m + \nu \nabla^2 \omega \quad (18)$$

with constant pressure and kinematic viscosity $\nu \equiv -i\hbar / 2m$. Thus, the weird quantum potential disappears completely and, hence, the Schrödinger equation reduces to classical diffusion and hydrodynamics but with complex transport coefficients. As is seen, the vacuum possesses purely imaginary diffusion and viscosity constants. They are complex-conjugated and their geometrically averaged value $\sqrt{\nu D} = \hbar / 2m$ equals to the Nelson universal diffusion constant [12]. Equations (17) and (18) open also a door to dissipative quantum mechanics [3] via including real parts of the complex transport coefficients as well.

The inconsistency of the Bohmian mechanics could be elucidated on the example of the classical Brownian motion, where the particles motion is described by the Langevin equation [13]

$$m\ddot{R} + b\dot{R} = -\nabla U + f_L \quad (19)$$

Here $b$ is the friction coefficient and $f_L$ is the stochastic Langevin force. Following Eq. (19) the probability density evolution is governed by two hydrodynamic-like equations [3]

$$\partial_t \rho = -\nabla \cdot (\rho V) \qquad m\partial_t V + mV \cdot \nabla V + bV = -\nabla(U + k_B T \ln \rho) \qquad (20)$$

Hence, according to thermodynamics the macroscopic image of the Langevin force is the gradient of the thermal free energy. Using Eq. (8) these equations can be easily transformed to

$$\partial_t \rho = -\nabla \cdot (\rho \nabla S / m) \qquad \partial_t S + (\nabla S)^2 / 2m + U + k_B T \ln \rho = -bS/m \qquad (21)$$

Following the Bohm logic, one can interpret the second equation as a dissipative thermal Hamilton-Jacobi equation, where the thermal chemical potential plays the role of $Q$. Employing now the de Broglie-Bohm guiding equation (5) one can derive a thermal Newtonian equation

$$m\ddot{R} + b\dot{R} = -\nabla(U + k_B T \ln \rho) \qquad (22)$$

being analogical of Eq. (6). Obviously Eq. (22) is not correct. It represents a mean-field deterministic approximation of the real stochastic Brownian dynamics rigorously described by Eq. (19). The main philosophical problem of the Bohmian mechanics is that it considers the quantum mechanics as a classical one, while it is, in fact, a kind of statistical mechanics.

Finally, in the case of the quantum Brownian motion both the quantum and Langevin stochastic forces are simultaneously acting and Eq. (19) advances to [3]

$$m\ddot{R} + b\dot{R} = -\nabla U + f_Q + f_L \qquad (23)$$

In the case of Bohmian approximation, one can replace the quantum stochastic force in Eq. (23) by the Bohm quantum force $-\nabla Q$ to obtain a density functional Bohm-Langevin equation [14]

$$m\ddot{R} + b\dot{R} = -\nabla(U + Q) + f_L \qquad (24)$$